# Three-Dimensional Generalized Dynamics of Soft-Matter Quasicrystals

Tian-You Fan[1]* and Zhi-Yi Tang [2]

1 School of Physics, Beijing Institute of Technology, Beijing 100081,China
2 School of Computer Science and Technology, Beijing Institute of Technology, Beijing 100081, China
*Corresponding author, e-mail:tyfan2013@163.com

**Abstract** The three-dimensional generalized dynamics of soft-matter quasicrystals was investigated, in which the governing equations of the dynamics are derived for observed 12-fold symmetry quasicrystals and possible observed 8- and 10-symmetry ones in near future in soft matter. The solving methods, possible solutions for some initial and boundary value problems of the equations and possible applications are discussed as well.

## 1. Introduction

Fan[1,2] discussed the soft-matter quasicrystals observed since 2004 [3-7],gave their dynamics equations, but only concerned the planar field. All observed soft-matter quasicrystals are two-dimensional quasicrystals so far. The two-dimensional quasicrystals consist of two kinds of ones, i.e., the first and second kinds. The first kind ones have been studied from the three-dimensional point of view of the group representation, while the second kind ones only have been studied from the two-dimensional point of view only. Therefore, the so-called three-dimensional analysis here gives only for the first kind ones. The three-dimensional field equations are very important and useful, for example, the flow of the soft-matter past a sphere, i.e., the generalized Stokes problem is significant in theory and application, it is well-known the classical Stokes solution was used by Einstein in 1905 in the Brownian motion and led to determination of Loschmidt's number, then again used by Millikan in 1922 led to the determination of the electronic charge. In soft matter many motion including the supramolecules self-assembly are related with the sphere configuration etc.

## 2. Generalized dynamics of soft-matter quasicrystals

In Refs [1,2], the author pointed out there were quite great differences between soft-matter and solid quasicrystals, from the point of view of condensed matter physics, but the formulation of the generalized hydrodynamics of the solid ones given by Lubensky et al [8,9] can be drawn for the application by the soft-matter ones, and after some modifications and supplementations, we can set up the generalized hydrodynamics or generalized dynamics for simplicity. The derivation of both dynamics is based on the generalized Langevin equation and the Poisson brackets, there are many similarities to each other. However there are main differences in principle between them:

1) The solid viscosity constitutive equation in Refs [8,9]

$$\sigma'_{ij} = \eta_{ijkl}\dot{\xi}_{kl}, \quad \dot{\xi}_{kl} = \frac{1}{2}\left(\frac{\partial V_k}{\partial x_l} + \frac{\partial V_l}{\partial x_k}\right)$$

is replaced by the fluid constitutive equation

$$p_{ij} = -p\delta_{ij} + \sigma'_{ij} = -p\delta_{ij} + \eta_{ijkl}\dot{\xi}_{kl}, \quad \sigma'_{ij} = \eta_{ijkl}\dot{\xi}_{kl} \quad \dot{\xi}_{kl} = \frac{1}{2}\left(\frac{\partial V_k}{\partial x_l} + \frac{\partial V_l}{\partial x_k}\right);$$

so that the constitutive laws of phonons, phasons and fluid phonon are, respectively,

$$\left.\begin{aligned}
&\sigma_{ij} = C_{ijkl}\varepsilon_{ij} + R_{ijkl}w_{kl}, \\
&H_{ij} = K_{ijkl}w_{ij} + R_{klij}\varepsilon_{kl}, \\
&p_{ij} = -p\delta_{ij} + \sigma'_{ij}, \sigma'_{ij} = \eta_{ijkl}\dot{\xi}_{kl}, \\
&\varepsilon_{ij} = \frac{1}{2}\left(\frac{\partial u_i}{\partial x_j} + \frac{\partial u_j}{\partial x_i}\right), w_{ij} = \frac{\partial w_i}{\partial x_j}, \dot{\xi}_{ij} = \frac{1}{2}\left(\frac{\partial V_i}{\partial x_j} + \frac{\partial V_j}{\partial x_i}\right)
\end{aligned}\right\} \quad (1)$$

where $u_i$ denotes phonon displacement vector; $\sigma_{ij}$ the phonon stress tensor; $\varepsilon_{ij}$ the phonon strain tensor; $w_i$ the phason displacement vector; $H_{ij}$ the phason stress tensor; $w_{ij}$ the phason strain tensor; $V_i$ the fluid phonon velocity vector; $p_{ij}$ the fluid stress tensor; $p$ the fluid pressure; $\eta_{ijkl}$ the fluid viscosity coefficient tensor; $\dot{\xi}_{ij}$ the fluid deformation rate tensor; and $C_{ijkl}, K_{ijkl}$ and $R_{ijkl}, R_{klij}$ are the phonon, phason and phonon-phason coupling elastic constant tensors, respectively. For simplicity, we here only discuss the constitutive law of simplest fluid, i.e.,

$$p_{ij} = -p\delta_{ij} + \sigma'_{ij} = -p\delta_{ij} + 2\eta(\dot{\xi}_{ij} - \frac{1}{3}\dot{\xi}_{kk}\delta_{ij}) + \eta'\dot{\xi}_{kk}\delta_{ij},$$

$$\sigma'_{ij} = 2\eta(\dot{\xi}_{ij} - \frac{1}{3}\dot{\xi}_{kk}\delta_{ij}) + \eta'\dot{\xi}_{kk}\delta_{ij},$$

$$\dot{\xi}_{kk} = \dot{\xi}_{11} + \dot{\xi}_{22} + \dot{\xi}_{33}, \dot{\xi}_{ij} = \frac{1}{2}\left(\frac{\partial V_i}{\partial x_j} + \frac{\partial V_j}{\partial x_i}\right)$$

where $\eta$ is the first viscosity coefficient; $\eta'$ the second one, which is omitted as it is too small(note that the description here only shows the difference of constitutive laws between solid and fluid, however this does not mean that in solid there is no pressure);

2) An equation of state $p = f(\rho)$ should be supplemented and in solid quasicrystals an equation is unnecessary. The equation of state belongs to the thermodynamics of soft matter, so the present discussion is beyond the scope of pure hydrodynamics. We used the results from Wensink [10], with some modifications by the author [1], i.e.,

$$p = f(\rho) = 3\frac{k_B T}{l^3 \rho_0^3}\left(\rho_0^2\rho + \rho_0\rho^2 + \rho^3\right) \quad (2)$$

where $k_B$ is the Boltzmann constant; $T$ the absolute temperature; $l$ the characteristic size of soft matter; and $\rho_0$ the initial value of mass density $\rho$.

After the modification and supplementation, we obtain the governing equations of hydrodynamics of soft-matter quasicrystals are as follows

$$\left.\begin{aligned}
&\frac{\partial \rho}{\partial t}+\nabla_k(\rho V_k)=0 \\
&\frac{\partial g_i(\mathrm{r},t)}{\partial t}=-\nabla_k(\mathrm{r})(V_k g_i)+\nabla_j(\mathrm{r})\left(-p\delta_{ij}+\eta_{ijkl}\nabla_k(\mathrm{r})V_l\right)-\left(\delta_{ij}-\nabla_i u_j\right)\frac{\delta H}{\delta u_j(\mathrm{r},t)} \\
&\quad +\left(\nabla_i w_j\right)\frac{\delta H}{\delta w_j(\mathrm{r},t)}-\rho\nabla_i(\mathrm{r})\frac{\delta H}{\delta\rho(\mathrm{r},t)},\qquad g_j=\rho V_j \\
&\frac{\partial u_i(\mathrm{r},t)}{\partial t}=-V_j\nabla_j(\mathrm{r})u_i-\Gamma_u\frac{\delta H}{\delta u_i(\mathrm{r},t)}+V_i \\
&\frac{\partial w_i(\mathrm{r},t)}{\partial t}=-V_j\nabla_j(\mathrm{r})w_i-\Gamma_w\frac{\delta H}{\delta w_i(\mathrm{r},t)} \\
&p=f(\rho)=3\frac{k_B T}{l^3\rho_0^3}\left(\rho_0^2\rho+\rho_0\rho^2+\rho^3\right)
\end{aligned}\right\} \quad (3)$$

in which $H$ is the energy functional or the Hamiltonians for quasicrystal systems. For the first kind of two-dimensional quasicrystals of soft matter, the energy functional or the Hamiltonians are similar to that given by Lubensky et al for solid quasicrystals in form such as:

$$H=H[\Psi(\mathrm{r},t)]=\int\frac{\mathrm{g}^2}{2\rho}d^d\mathrm{r}+\int\left[\frac{1}{2}A\left(\frac{\delta\rho}{\rho_0}\right)^2+B\left(\frac{\delta\rho}{\rho_0}\right)\nabla\bullet\mathrm{u}\right]d^d\mathrm{r}+F_{el}$$

$$=H_{kin}+H_{density}+F_{el}$$

$$\mathrm{g}=\rho\mathrm{V}, F_{el}=F_u+F_w+F_{uw} \qquad (4)$$

where $\delta\rho=\rho-\rho_0$, $A$, $B$ the material constants due the variation of the density, and $F_{el}$ denotes the elastic strain energy; and $F_u, F_w, F_{uw}$ represent the strain energies of phonons, phasons and phonon-phason coupling of the matter, respectively:

$$\begin{aligned}
F_u &= \int\frac{1}{2}C_{ijkl}\varepsilon_{ij}\varepsilon_{kl}d^d\mathrm{r} \\
F_w &= \int\frac{1}{2}K_{ijkl}w_{ij}w_{kl}d^d\mathrm{r} \\
F_{uw} &= \int\left(R_{ijkl}\varepsilon_{ij}w_{kl}+R_{klij}w_{ij}\varepsilon_{kl}\right)d^d\mathrm{r}
\end{aligned} \qquad (5)$$

The derivation of the first 4 equations (the equations of motion) is based on the Poisson bracket method of condensed matter physics [11], which was used by Lubensky et al. [8] in solid quasicrystal study for the first time. The Chinese literature on this method can be found in Ref [1] and others provided by the author, where there are many additional details on the derivations of equations of motion of quasicrystals. A key application of the Poisson bracket method lies in the Hamiltonian individual quasicrystal systems, given in the following sections. It is evident that the derivation based on Poisson brackets are not concerned the equation of state, which is a result of thermodynamics [10].

The modified framework of hydrodynamics described by equations (3) is called as generalized hydrodynamics, which is a heritage and development of equations of Lubensky et al created for solid quasicrystals.

## 3. Soft-matter quasicrystals with 12-fold symmetry

The discussion in Ref [1] was only concerned with the planar field, we there did not concern the three-dimensional problem of the dynamics, and equations (3) here hold for one-, two- and three-dimensions. The equations (3) are tight, but very hard to understand, and it is not convenient to apply in the calculation.

At first we simplify the equations for soft-matter quasicrystals with 12-fold symmetry, which might be the most important class of soft-matter quasicrystals observed so far. For this purpose, we must list the three-dimensional constitutive laws on phonons, phasons and fluid phonon, respectively, as follows Ref [12-14]

$$\left.\begin{aligned}
\sigma_{xx} &= C_{11}\varepsilon_{xx} + C_{12}\varepsilon_{yy} + C_{13}\varepsilon_{zz} \\
\sigma_{yy} &= C_{12}\varepsilon_{xx} + C_{11}\varepsilon_{yy} + C_{13}\varepsilon_{zz} \\
\sigma_{zz} &= C_{13}\varepsilon_{xx} + C_{13}\varepsilon_{yy} + C_{33}\varepsilon_{zz} \\
\sigma_{yz} &= \sigma_{zy} = 2C_{44}\varepsilon_{yz} \\
\sigma_{zx} &= \sigma_{xz} = 2C_{44}\varepsilon_{zx} \\
\sigma_{xy} &= \sigma_{yx} = 2C_{66}\varepsilon_{xy} \\
H_{xx} &= K_1 w_{xx} + K_2 w_{yy} \\
H_{yy} &= K_2 w_{xx} + K_1 w_{yy} \\
H_{yz} &= K_4 w_{yz} \\
H_{xy} &= (K_1 + K_2 + K_3) w_{xy} + K_3 w_{yx} \\
H_{xz} &= K_4 w_{xz} \\
H_{yx} &= K_3 w_{xy} + (K_1 + K_2 + K_3) w_{yx} \\
p_{xx} &= -p + 2\eta\dot{\xi}_{xx} - \frac{2}{3}\eta\dot{\xi}_{kk} \\
p_{yy} &= -p + 2\eta\dot{\xi}_{yy} - \frac{2}{3}\eta\dot{\xi}_{kk} \\
p_{zz} &= -p + 2\eta\dot{\xi}_{zz} - \frac{2}{3}\eta\dot{\xi}_{kk} \\
p_{yz} &= 2\eta\dot{\xi}_{yz} \\
p_{zx} &= 2\eta\dot{\xi}_{zx} \\
p_{xy} &= 2\eta\dot{\xi}_{xy}
\end{aligned}\right\} \quad (6)$$

then the equations of dynamics of soft-matter quasicrystals of 12-fold symmetry are as following by omitting the higher order terms of $\nabla_i\left(u_j \frac{\delta H}{\delta u_j}\right)$ and $\nabla_i\left(w_j \frac{\delta H}{\delta w_j}\right)$ listed in equations (3)

$$\left.\begin{aligned}
&\frac{\partial \rho}{\partial t}+\nabla\cdot(\rho\mathbf{V})=0 \\
&\frac{\partial(\rho V_x)}{\partial t}+\frac{\partial(V_x\rho V_x)}{\partial x}+\frac{\partial(V_y\rho V_x)}{\partial y}+\frac{\partial(V_z\rho V_x)}{\partial z}=-\frac{\partial p}{\partial x}+\eta\nabla^2 V_x+\frac{1}{3}\eta\frac{\partial}{\partial x}\nabla\cdot\mathbf{V}\\
&+\left(C_{11}\frac{\partial^2}{\partial x^2}+C_{66}\frac{\partial^2}{\partial y^2}+C_{44}\frac{\partial^2}{\partial z^2}\right)u_x+(C_{12}+C_{66})\frac{\partial^2 u_y}{\partial x\partial y}+(C_{13}+C_{44})\frac{\partial^2 u_z}{\partial x\partial z}-B\frac{\partial}{\partial x}\nabla\cdot\mathbf{u}-(A-B)\frac{1}{\rho_0}\frac{\partial\delta\rho}{\partial x}\\
&\frac{\partial(\rho V_y)}{\partial t}+\frac{\partial(V_x\rho V_y)}{\partial x}+\frac{\partial(V_y\rho V_y)}{\partial y}+\frac{\partial(V_z\rho V_y)}{\partial z}=-\frac{\partial p}{\partial y}+\eta\nabla^2 V_y+\frac{1}{3}\eta\frac{\partial}{\partial y}\nabla\cdot\mathbf{V}\\
&+(C_{12}+C_{66})\frac{\partial^2 u_x}{\partial x\partial y}+\left(C_{66}\frac{\partial^2}{\partial x^2}+C_{11}\frac{\partial^2}{\partial y^2}+C_{44}\frac{\partial^2}{\partial z^2}\right)u_y+(C_{13}+C_{44})\frac{\partial^2 u_z}{\partial y\partial z}-B\frac{\partial}{\partial y}\nabla\cdot\mathbf{u}-(A-B)\frac{1}{\rho_0}\frac{\partial\delta\rho}{\partial y}\\
&\frac{\partial(\rho V_z)}{\partial t}+\frac{\partial(V_x\rho V_z)}{\partial x}+\frac{\partial(V_y\rho V_z)}{\partial y}+\frac{\partial(V_z\rho V_z)}{\partial z}=-\frac{\partial p}{\partial z}+\eta\nabla^2 V_z+\frac{1}{3}\eta\frac{\partial}{\partial z}\nabla\cdot\mathbf{V}\\
&+(C_{13}+C_{44})\left(\frac{\partial^2 u_x}{\partial x\partial z}+\frac{\partial^2 u_y}{\partial y\partial z}\right)+\left(C_{44}\frac{\partial^2}{\partial x^2}+C_{44}\frac{\partial^2}{\partial y^2}+C_{33}\frac{\partial^2}{\partial z^2}\right)u_z-B\frac{\partial}{\partial z}\nabla\cdot\mathbf{u}-(A-B)\frac{1}{\rho_0}\frac{\partial\delta\rho}{\partial z}\\
&\frac{\partial u_x}{\partial t}+V_x\frac{\partial u_x}{\partial x}+V_y\frac{\partial u_x}{\partial y}+V_z\frac{\partial u_x}{\partial z}=V_x\\
&\quad+\Gamma_{\mathbf{u}}\left[\left(C_{11}\frac{\partial^2}{\partial x^2}+C_{66}\frac{\partial^2}{\partial y^2}+C_{44}\frac{\partial^2}{\partial z^2}\right)u_x+(C_{12}+C_{66})\frac{\partial^2 u_y}{\partial x\partial y}+(C_{13}+C_{44})\frac{\partial^2 u_z}{\partial x\partial z}\right]\\
&\frac{\partial u_y}{\partial t}+V_x\frac{\partial u_y}{\partial x}+V_y\frac{\partial u_y}{\partial y}+V_z\frac{\partial u_y}{\partial z}=V_y\\
&\quad+\Gamma_{\mathbf{u}}\left[(C_{12}+C_{66})\frac{\partial^2 u_x}{\partial x\partial y}+\left(C_{66}\frac{\partial^2}{\partial x^2}+C_{11}\frac{\partial^2}{\partial y^2}+C_{44}\frac{\partial^2}{\partial z^2}\right)u_y+(C_{13}+C_{44})\frac{\partial^2 u_z}{\partial y\partial z}\right]\\
&\frac{\partial u_z}{\partial t}+V_x\frac{\partial u_z}{\partial x}+V_y\frac{\partial u_z}{\partial y}+V_z\frac{\partial u_z}{\partial z}=V_z\\
&\quad+\Gamma_{\mathbf{u}}\left[(C_{13}+C_{44})\left(\frac{\partial^2 u_x}{\partial x\partial z}+\frac{\partial^2 u_y}{\partial y\partial z}\right)+\left(C_{44}\frac{\partial^2}{\partial x^2}+C_{44}\frac{\partial^2}{\partial y^2}+C_{33}\frac{\partial^2}{\partial z^2}\right)u_z\right]\\
&\frac{\partial w_x}{\partial t}+V_x\frac{\partial w_x}{\partial x}+V_y\frac{\partial w_x}{\partial y}+V_z\frac{\partial w_x}{\partial z}=\Gamma_{\mathbf{w}}\left[\left(K_1\frac{\partial^2}{\partial x^2}+(K_1+K_2+K_3)\frac{\partial^2}{\partial y^2}+K_4\frac{\partial^2}{\partial z^2}\right)w_x+(K_2+K_3)\frac{\partial^2 w_y}{\partial x\partial y}\right]\\
&\frac{\partial w_y}{\partial t}+V_x\frac{\partial w_y}{\partial x}+V_y\frac{\partial w_y}{\partial y}+V_z\frac{\partial w_y}{\partial z}=\Gamma_{\mathbf{w}}\left[(K_2+K_3)\frac{\partial^2 w_x}{\partial x\partial y}+\left((K_1+K_2+K_3)\frac{\partial^2}{\partial x^2}+K_1\frac{\partial^2}{\partial y^2}+K_4\frac{\partial^2}{\partial z^2}\right)w_y\right]\\
&p=f(\rho)=3\frac{k_B T}{l^3\rho_0^3}\left(\rho_0^2\rho+\rho_0\rho^2+\rho^3\right)
\end{aligned}\right\} (7)$$

in which $\nabla^2=\frac{\partial^2}{\partial x^2}+\frac{\partial^2}{\partial y^2}+\frac{\partial^2}{\partial z^2}$, $\nabla=\mathbf{i}\frac{\partial}{\partial x}+\mathbf{j}\frac{\partial}{\partial y}+\mathbf{k}\frac{\partial}{\partial z}$, $\mathbf{V}=\mathbf{i}V_x+\mathbf{j}V_y+\mathbf{k}V_z$, $\mathbf{u}=\mathbf{i}u_x+\mathbf{j}u_y+\mathbf{k}u_z$, and $C_{11},C_{12},C_{13},C_{33},C_{44},C_{66}=(C_{11}-C_{12})/2$ the phonon elastic constants, $K_1,K_2,K_3,K_4$ the phason elastic constants (refer to[1], $\eta$ the fluid dynamic viscosity, and $\Gamma_u,\Gamma_w$ the phonon and phason dissipation coefficients, $A,B$ the material constants due to variation of mass density, respectively.

The equations (7) are the final governing equations of dynamics of soft-matter quasicrystals of 12-fold symmetry in three-dimensional case with fields variables $u_x, u_y, u_z, w_x, w_y, V_x, V_y, V_z, \rho$ and $p$, the amount of the field variables is 10, and amount of field equations is 10 too, among them: (7a) is the mass conservation equation, (7b)-(7d) the momentum conservation equations, (7e)- (7g) the equations of motion of phonons due to the symmetry breaking, (7h) and (7i) the phason dissipation equations, and (7j) the equation of state, respectively. The equations are consistent to be mathematical solvability, if there is lack of the equation of state, the equation system is not closed, and has no meaning mathematically and physically. This shows the equation of state is necessary.

These equations reveal the nature of wave propagation of fields $\mathbf{u}$ and $\mathbf{V}$ with phonon wave speeds $c_1 = \sqrt{\dfrac{2A + C_{11} - B}{\rho}}$, $c_2 = c_3 = \sqrt{\dfrac{C_{11} - C_{12}}{2\rho}}$ and fluid phonon wave speed $c_4 = \sqrt{\left(\dfrac{\partial p}{\partial \rho}\right)_s}$ and the nature of the diffusion of field $\mathbf{w}$ with diffusive coefficient $D = \dfrac{1}{\Gamma_w}$ from the view point of hydrodynamics.

## 4. Soft-matter quasicrystals with 8-fold symmetry

Apart from the observed 12- and 18-fold symmetry soft-matter quasicrystals, the 8-fold symmetrical soft-matter quasicrystals may also be observed in the near future. This kind of solid quasicrystal is very stable, which is important especially as there are strong coupling effects between the phonons and phasons, and it is interesting to study their mechanical and physical properties and mathematical solutions. We considered the plane of quasiperiodicity to be $xy-$plane, if the $z-$axis is 8-fold symmetry axis. Next, for the possibility of soft-matter octagonal quasicrystals in soft matter there is the final governing equation system of the generalized dynamics, after some derivations by similar steps to the previous section, but we must list the constitutive law [12-14] first:

$$\left.\begin{aligned}
\sigma_{xx} &= C_{11}\varepsilon_{xx} + C_{12}\varepsilon_{yy} + C_{13}\varepsilon_{zz} + R(w_{xx} + w_{yy}) \\
\sigma_{yy} &= C_{12}\varepsilon_{xx} + C_{11}\varepsilon_{yy} + C_{13}\varepsilon_{zz} - R(w_{xx} + w_{yy}) \\
\sigma_{zz} &= C_{13}\varepsilon_{xx} + C_{13}\varepsilon_{yy} + C_{33}\varepsilon_{zz} \\
\sigma_{yz} &= \sigma_{zy} = 2C_{44}\varepsilon_{yz} \\
\sigma_{zx} &= \sigma_{xz} = 2C_{44}\varepsilon_{zx} \\
\sigma_{xy} &= \sigma_{yx} = 2C_{66}\varepsilon_{xy} - Rw_{xy} + Rw_{yx} \\
H_{xx} &= K_1 w_{xx} + K_2 w_{yy} + R(\varepsilon_{xx} - \varepsilon_{yy}) \\
H_{yy} &= K_2 w_{xx} + K_1 w_{yy} + R(\varepsilon_{xx} - \varepsilon_{yy}) \\
H_{yz} &= K_4 w_{yz} \\
H_{xy} &= (K_1 + K_2 + K_3) w_{xy} + K_3 w_{yx} - 2R\varepsilon_{xy} \\
H_{xz} &= K_4 w_{xz} \\
H_{yx} &= K_3 w_{xy} + (K_1 + K_2 + K_3) w_{yx} + 2R\varepsilon_{xy} \\
p_{xx} &= -p + 2\eta \dot{\xi}_{xx} - \frac{2}{3}\eta \dot{\xi}_{kk} \\
p_{yy} &= -p + 2\eta \dot{\xi}_{yy} - \frac{2}{3}\eta \dot{\xi}_{kk} \\
p_{zz} &= -p + 2\eta \dot{\xi}_{zz} - \frac{2}{3}\eta \dot{\xi}_{kk} \\
p_{yz} &= 2\eta \dot{\xi}_{yz} \\
p_{zx} &= 2\eta \dot{\xi}_{zx} \\
p_{xy} &= 2\eta \dot{\xi}_{xy}
\end{aligned}\right\} \quad (8)$$

With this basic relations the governing equations of three-dimensional generalized dynamics of soft-matter quasicrystals with 8-fold symmetry are as follows

$$\frac{\partial \rho}{\partial t} + \nabla \cdot (\rho \mathbf{V}) = 0$$

$$\frac{\partial(\rho V_x)}{\partial t} + \frac{\partial(V_x \rho V_x)}{\partial x} + \frac{\partial(V_y \rho V_x)}{\partial y} + \frac{\partial(V_z \rho V_x)}{\partial z} = -\frac{\partial p}{\partial x} + \eta \nabla^2 V_x + \frac{1}{3}\eta \frac{\partial}{\partial x}\nabla \cdot \mathbf{V}$$

$$+ \left(C_{11}\frac{\partial^2}{\partial x^2} + C_{66}\frac{\partial^2}{\partial y^2} + C_{44}\frac{\partial^2}{\partial z^2}\right)u_x + (C_{12}+C_{66})\frac{\partial^2 u_y}{\partial x \partial y} + (C_{13}+C_{44})\frac{\partial^2 u_z}{\partial x \partial z} + R\left(\frac{\partial^2}{\partial x^2} - \frac{\partial^2}{\partial y^2}\right)w_x$$

$$- B\frac{\partial}{\partial x}\nabla \cdot \mathbf{u} - (A-B)\frac{1}{\rho_0}\frac{\partial \delta\rho}{\partial x}$$

$$\frac{\partial(\rho V_y)}{\partial t} + \frac{\partial(V_x \rho V_y)}{\partial x} + \frac{\partial(V_y \rho V_y)}{\partial y} + \frac{\partial(V_z \rho V_y)}{\partial z} = -\frac{\partial p}{\partial y} + \eta \nabla^2 V_y + \frac{1}{3}\eta \frac{\partial}{\partial y}\nabla \cdot \mathbf{V}$$

$$+ (C_{12}+C_{66})\frac{\partial^2 u_x}{\partial x \partial y} + \left(C_{66}\frac{\partial^2}{\partial x^2} + C_{11}\frac{\partial^2}{\partial y^2} + C_{44}\frac{\partial^2}{\partial z^2}\right)u_y + (C_{13}+C_{44})\frac{\partial^2 u_z}{\partial y \partial z} - R\left(\frac{\partial^2 w_x}{\partial x \partial y} + \frac{\partial^2 w_y}{\partial y^2}\right)$$

$$- B\frac{\partial}{\partial y}\nabla \cdot \mathbf{u} - (A-B)\frac{1}{\rho_0}\frac{\partial \delta\rho}{\partial y}$$

$$\frac{\partial(\rho V_z)}{\partial t} + \frac{\partial(V_x \rho V_z)}{\partial x} + \frac{\partial(V_y \rho V_z)}{\partial y} + \frac{\partial(V_z \rho V_z)}{\partial z} = -\frac{\partial p}{\partial z} + \eta \nabla^2 V_z + \frac{1}{3}\eta \frac{\partial}{\partial z}\nabla \cdot \mathbf{V}$$

$$+ (C_{13}+C_{44})\left(\frac{\partial^2 u_x}{\partial x \partial z} + \frac{\partial^2 u_y}{\partial y \partial z}\right) + \left(C_{44}\frac{\partial^2}{\partial x^2} + C_{44}\frac{\partial^2}{\partial y^2} + C_{33}\frac{\partial^2}{\partial z^2}\right)u_z - B\frac{\partial}{\partial z}\nabla \cdot \mathbf{u} - (A-B)\frac{1}{\rho_0}\frac{\partial \delta\rho}{\partial z}$$

$$\frac{\partial u_x}{\partial t} + V_x \frac{\partial u_x}{\partial x} + V_y \frac{\partial u_x}{\partial y} + V_z \frac{\partial u_x}{\partial z} = V_x$$

$$+ \Gamma_{\mathbf{u}}\left[\left(C_{11}\frac{\partial^2}{\partial x^2} + C_{66}\frac{\partial^2}{\partial y^2} + C_{44}\frac{\partial^2}{\partial z^2}\right)u_x + (C_{12}+C_{66})\frac{\partial^2 u_y}{\partial x \partial y} + (C_{13}+C_{44})\frac{\partial^2 u_z}{\partial x \partial z} + R\left(\frac{\partial^2}{\partial x^2} - \frac{\partial^2}{\partial y^2}\right)w_x\right]$$

$$\frac{\partial u_y}{\partial t} + V_x \frac{\partial u_y}{\partial x} + V_y \frac{\partial u_y}{\partial y} + V_z \frac{\partial u_y}{\partial z} = V_y$$

$$+ \Gamma_{\mathbf{u}}\left[(C_{12}+C_{66})\frac{\partial^2 u_x}{\partial x \partial y} + \left(C_{66}\frac{\partial^2}{\partial x^2} + C_{11}\frac{\partial^2}{\partial y^2} + C_{44}\frac{\partial^2}{\partial z^2}\right)u_y + (C_{13}+C_{44})\frac{\partial^2 u_z}{\partial y \partial z} - R\left(\frac{\partial^2 w_x}{\partial x \partial y} + \frac{\partial^2 w_y}{\partial y^2}\right)\right]$$

$$\frac{\partial u_z}{\partial t} + V_x \frac{\partial u_z}{\partial x} + V_y \frac{\partial u_z}{\partial y} + V_z \frac{\partial u_z}{\partial z} = V_z$$

$$+ \Gamma_{\mathbf{u}}\left[(C_{13}+C_{44})\left(\frac{\partial^2 u_x}{\partial x \partial z} + \frac{\partial^2 u_y}{\partial y \partial z}\right) + \left(C_{44}\frac{\partial^2}{\partial x^2} + C_{44}\frac{\partial^2}{\partial y^2} + C_{33}\frac{\partial^2}{\partial z^2}\right)u_z\right]$$

$$\frac{\partial w_x}{\partial t} + V_x \frac{\partial w_x}{\partial x} + V_y \frac{\partial w_x}{\partial y} + V_z \frac{\partial w_x}{\partial z} = \Gamma_{\mathbf{w}}\left[\left(K_1 \frac{\partial^2}{\partial x^2} + (K_1+K_2+K_3)\frac{\partial^2}{\partial y^2} + K_4 \frac{\partial^2}{\partial z^2}\right)w_x + (K_2+K_3)\frac{\partial^2 w_y}{\partial x \partial y}\right.$$

$$\left. + R\left(\frac{\partial^2}{\partial x^2} - \frac{\partial^2}{\partial y^2}\right)u_x - 2R\frac{\partial^2 u_y}{\partial x \partial y}\right]$$

$$\frac{\partial w_y}{\partial t} + V_x \frac{\partial w_y}{\partial x} + V_y \frac{\partial w_y}{\partial y} V_z \frac{\partial w_y}{\partial z} = \Gamma_{\mathbf{w}}\left[(K_2+K_3)\frac{\partial^2 w_x}{\partial x \partial y} + \left((K_1+K_2+K_3)\frac{\partial^2}{\partial x^2} + K_1 \frac{\partial^2}{\partial y^2} + K_4 \frac{\partial^2}{\partial z^2}\right)w_y\right.$$

$$\left. + 2R\frac{\partial^2 u_x}{\partial x \partial y} + R\left(\frac{\partial^2}{\partial x^2} - \frac{\partial^2}{\partial y^2}\right)u_y\right]$$

$$p = f(\rho) = 3\frac{k_B T}{l^3 \rho_0^3}\left(\rho_0^2 \rho + \rho_0 \rho^2 + \rho^3\right)$$

$$(9)$$

in which $\nabla^2 = \dfrac{\partial^2}{\partial x^2} + \dfrac{\partial^2}{\partial y^2} + \dfrac{\partial^2}{\partial z^2}$, $\nabla = \mathbf{i}\dfrac{\partial}{\partial x} + \mathbf{j}\dfrac{\partial}{\partial y} + \mathbf{k}\dfrac{\partial}{\partial z}$, $\mathbf{V} = \mathbf{i}V_x + \mathbf{j}V_y + \mathbf{k}V_z$,

$\mathbf{u} = \mathbf{i}u_x + \mathbf{j}u_y + \mathbf{k}u_z$, $\mathbf{w} = \mathbf{i}w_x + \mathbf{j}w_y$ and $C_{11}, C_{12}, C_{13}, C_{33}, C_{44}, C_{66} = (C_{11} - C_{12})/2$ the phonon elastic constants, $K_1, K_2, K_3, K_4$ the phason elastic constants, $R$ the phonon-phason coupling constant, $\eta$ the fluid dynamic viscosity, and $\Gamma_u, \Gamma_w$ the phonon and phason dissipation coefficients, $A, B$ the material constants due to variation of mass density, respectively.

The equations (9) are the final governing equations of dynamics of soft-matter quasicrystals of 8-fold symmetry in three-dimensional case with fields variables $u_x, u_y, u_z, w_x, w_y, V_x, V_y, V_z, \rho$ and $p$, the amount of the field variables is 10, and amount of field equations is 10 too, among them: (9a) is the mass conservation equation, (9b)-(9d) the momentum conservation equations, (9e)-(9g) the equations of motion of phonons due to the symmetry breaking, (9h) and (9i) the phason dissipation equations, and (9j) the equation of state, respectively. The equations are consistent to be mathematical solvability, if there is lack of the equation of state, the equation system is not closed, and has no meaning mathematically and physically. This shows the equation of state is necessary.

These equations reveal the nature of wave propagation of fields $\mathbf{u}$ and $\mathbf{V}$ with phonon wave speeds $c_1 = \sqrt{\dfrac{2A + C_{11} - B}{\rho}}$, $c_2 = c_3 = \sqrt{\dfrac{C_{11} - C_{12}}{2\rho}}$ and fluid phonon wave speed $c_4 = \sqrt{\left(\dfrac{\partial p}{\partial \rho}\right)_s}$ and the nature of the diffusion of field $\mathbf{w}$ with diffusive coefficient $D = \dfrac{1}{\Gamma_w}$ from the view point of hydrodynamics.

## 5. Soft-matter quasicrystals with 10-fold symmetry

The 12- and 18-fold symmetrical soft-matter quasicrystals were observed, with a possibility that 10-fold symmetrical soft-matter quasicrystals will be discovered thereafter. These kind of solid quasicrystals are very stable, which promote important interest. Especially as there are strong coupling effects between the phonons and phasons, it is interesting to study their mechanical and physical properties and mathematical solutions. If we consider the plane in the $xy-$plane to be a quasiperiodic plane, and if $z-$axis is the 10-fold symmetry axis, then after derivation similar to those previous sections, we can obtain the final governing equation system for the type of soft-matter quasicrystals, but we must list the constitutive laws for phonons, phasons and fluid phonon[12-14] first:

$$\left.\begin{aligned}
\sigma_{xx} &= C_{11}\varepsilon_{xx} + C_{12}\varepsilon_{yy} + C_{13}\varepsilon_{zz} + R(w_{xx} + w_{yy}) \\
\sigma_{yy} &= C_{12}\varepsilon_{xx} + C_{11}\varepsilon_{yy} + C_{13}\varepsilon_{zz} - R(w_{xx} + w_{yy}) \\
\sigma_{zz} &= C_{13}\varepsilon_{xx} + C_{13}\varepsilon_{yy} + C_{33}\varepsilon_{zz} \\
\sigma_{yz} &= \sigma_{zy} = 2C_{44}\varepsilon_{yz} \\
\sigma_{zx} &= \sigma_{xz} = 2C_{44}\varepsilon_{zx} \\
\sigma_{xy} &= \sigma_{yx} = 2C_{66}\varepsilon_{xy} - R(w_{xy} - w_{yx}) \\
H_{xx} &= K_1 w_{xx} + K_2 w_{yy} + R(\varepsilon_{xx} - \varepsilon_{yy}) \\
H_{yy} &= K_2 w_{xx} + K_1 w_{yy} + R(\varepsilon_{xx} - \varepsilon_{yy}) \\
H_{yz} &= K_4 w_{yz} \\
H_{xy} &= K_1 w_{xy} - K_2 w_{yx} - 2R\varepsilon_{xy} \\
H_{xz} &= K_4 w_{xz} \\
H_{yx} &= -K_2 w_{xy} + K_1 w_{yx} + 2R\varepsilon_{xy} \\
p_{xx} &= -p + 2\eta\dot{\xi}_{xx} - \frac{2}{3}\eta\dot{\xi}_{kk} \\
p_{yy} &= -p + 2\eta\dot{\xi}_{yy} - \frac{2}{3}\eta\dot{\xi}_{kk} \\
p_{zz} &= -p + 2\eta\dot{\xi}_{zz} - \frac{2}{3}\eta\dot{\xi}_{kk} \\
p_{yz} &= 2\eta\dot{\xi}_{yz} \\
p_{zx} &= 2\eta\dot{\xi}_{zx} \\
p_{xy} &= 2\eta\dot{\xi}_{xy}
\end{aligned}\right\} \quad (10)$$

So that we obtain the governing equations of 10-fold symmetry quasicrystals in soft matter

$$\left.\begin{aligned}
&\frac{\partial \rho}{\partial t}+\nabla\cdot(\rho\mathbf{V})=0 \\
&\frac{\partial(\rho V_x)}{\partial t}+\frac{\partial(V_x\rho V_x)}{\partial x}+\frac{\partial(V_y\rho V_x)}{\partial y}+\frac{\partial(V_z\rho V_x)}{\partial z}=-\frac{\partial p}{\partial x}+\eta\nabla^2 V_x+\frac{1}{3}\eta\frac{\partial}{\partial x}\nabla\cdot\mathbf{V} \\
&\quad+\left(C_{11}\frac{\partial^2}{\partial x^2}+C_{66}\frac{\partial^2}{\partial y^2}+C_{44}\frac{\partial^2}{\partial z^2}\right)u_x+(C_{12}+C_{66})\frac{\partial^2 u_y}{\partial x\partial y}+(C_{13}+C_{44})\frac{\partial^2 u_z}{\partial x\partial z}+R\left(\frac{\partial^2}{\partial x^2}-\frac{\partial^2}{\partial y^2}\right)w_x \\
&\quad-B\frac{\partial}{\partial x}\nabla\cdot\mathbf{u}-(A-B)\frac{1}{\rho_0}\frac{\partial\delta\rho}{\partial x} \\
&\frac{\partial(\rho V_y)}{\partial t}+\frac{\partial(V_x\rho V_y)}{\partial x}+\frac{\partial(V_y\rho V_y)}{\partial y}+\frac{\partial(V_z\rho V_y)}{\partial z}=-\frac{\partial p}{\partial y}+\eta\nabla^2 V_y+\frac{1}{3}\eta\frac{\partial}{\partial y}\nabla\cdot\mathbf{V} \\
&\quad+(C_{12}+C_{66})\frac{\partial^2 u_x}{\partial x\partial y}+\left(C_{66}\frac{\partial^2}{\partial x^2}+C_{11}\frac{\partial^2}{\partial y^2}+C_{44}\frac{\partial^2}{\partial z^2}\right)u_y+(C_{13}+C_{44})\frac{\partial^2 u_z}{\partial y\partial z}-R\left(\frac{\partial^2 w_x}{\partial x\partial y}+\frac{\partial^2 w_y}{\partial y^2}\right) \\
&\quad-B\frac{\partial}{\partial y}\nabla\cdot\mathbf{u}-(A-B)\frac{1}{\rho_0}\frac{\partial\delta\rho}{\partial y} \\
&\frac{\partial(\rho V_z)}{\partial t}+\frac{\partial(V_x\rho V_z)}{\partial x}+\frac{\partial(V_y\rho V_z)}{\partial y}+\frac{\partial(V_z\rho V_z)}{\partial z}=-\frac{\partial p}{\partial z}+\eta\nabla^2 V_z+\frac{1}{3}\eta\frac{\partial}{\partial z}\nabla\cdot\mathbf{V} \\
&\quad+(C_{13}+C_{44})\left(\frac{\partial^2 u_x}{\partial x\partial z}+\frac{\partial^2 u_y}{\partial y\partial z}\right)+\left(C_{44}\frac{\partial^2}{\partial x^2}+C_{44}\frac{\partial^2}{\partial y^2}+C_{33}\frac{\partial^2}{\partial z^2}\right)u_z-B\frac{\partial}{\partial z}\nabla\cdot\mathbf{u}-(A-B)\frac{1}{\rho_0}\frac{\partial\delta\rho}{\partial z} \\
&\frac{\partial u_x}{\partial t}+V_x\frac{\partial u_x}{\partial x}+V_y\frac{\partial u_x}{\partial y}+V_z\frac{\partial u_x}{\partial z}=V_x \\
&\quad+\Gamma_{\mathbf{u}}\left[\left(C_{11}\frac{\partial^2}{\partial x^2}+C_{66}\frac{\partial^2}{\partial y^2}+C_{44}\frac{\partial^2}{\partial z^2}\right)u_x+(C_{12}+C_{66})\frac{\partial^2 u_y}{\partial x\partial y}+(C_{13}+C_{44})\frac{\partial^2 u_z}{\partial x\partial z}+R\left(\frac{\partial^2}{\partial x^2}-\frac{\partial^2}{\partial y^2}\right)w_x\right] \\
&\frac{\partial u_y}{\partial t}+V_x\frac{\partial u_y}{\partial x}+V_y\frac{\partial u_y}{\partial y}+V_z\frac{\partial u_y}{\partial z}=V_y \\
&\quad+\Gamma_{\mathbf{u}}\left[(C_{12}+C_{66})\frac{\partial^2 u_x}{\partial x\partial y}+\left(C_{66}\frac{\partial^2}{\partial x^2}+C_{11}\frac{\partial^2}{\partial y^2}+C_{44}\frac{\partial^2}{\partial z^2}\right)u_y+(C_{13}+C_{44})\frac{\partial^2 u_z}{\partial y\partial z}-R\left(\frac{\partial^2 w_x}{\partial x\partial y}+\frac{\partial^2 w_y}{\partial y^2}\right)\right] \\
&\frac{\partial u_z}{\partial t}+V_x\frac{\partial u_z}{\partial x}+V_y\frac{\partial u_z}{\partial y}+V_z\frac{\partial u_z}{\partial z}=V_z \\
&\quad+\Gamma_{\mathbf{u}}\left[(C_{13}+C_{44})\left(\frac{\partial^2 u_x}{\partial x\partial z}+\frac{\partial^2 u_y}{\partial y\partial z}\right)+\left(C_{44}\frac{\partial^2}{\partial x^2}+C_{44}\frac{\partial^2}{\partial y^2}+C_{33}\frac{\partial^2}{\partial z^2}\right)u_z\right] \\
&\frac{\partial w_x}{\partial t}+V_x\frac{\partial w_x}{\partial x}+V_y\frac{\partial w_x}{\partial y}+V_z\frac{\partial w_x}{\partial z}=\Gamma_{\mathbf{w}}\left[\left(K_1\frac{\partial^2}{\partial x^2}+K_2\frac{\partial^2}{\partial x\partial y}+K_4\frac{\partial^2}{\partial z^2}\right)w_x+\left(K_1\frac{\partial^2}{\partial y^2}+K_2\frac{\partial^2}{\partial x\partial y}\right)w_y\right. \\
&\quad\left.+R\left(\frac{\partial^2}{\partial x^2}-\frac{\partial^2}{\partial y^2}\right)u_x-2R\frac{\partial^2 u_y}{\partial x\partial y}\right] \\
&\frac{\partial w_y}{\partial t}+V_x\frac{\partial w_y}{\partial x}+V_y\frac{\partial w_y}{\partial y}+V_z\frac{\partial w_y}{\partial z}=\Gamma_{\mathbf{w}}\left[\left(K_1\frac{\partial^2}{\partial x^2}+K_1\frac{\partial^2}{\partial x\partial y}+K_4\frac{\partial^2}{\partial z^2}\right)w_y\right. \\
&\quad\left.+2R\frac{\partial^2 u_x}{\partial x\partial y}+R\left(\frac{\partial^2}{\partial x^2}-\frac{\partial^2}{\partial y^2}\right)u_y\right] \\
&p=f(\rho)=3\frac{k_B T}{l^3 \rho_0^3}\left(\rho_0^2\rho+\rho_0\rho^2+\rho^3\right)
\end{aligned}\right\} \quad (11)$$

in which $\nabla^2 = \frac{\partial^2}{\partial x^2} + \frac{\partial^2}{\partial y^2} + \frac{\partial^2}{\partial z^2}$, $\nabla = \mathbf{i}\frac{\partial}{\partial x} + \mathbf{j}\frac{\partial}{\partial y} + \mathbf{k}\frac{\partial}{\partial z}$, $\mathbf{V} = \mathbf{i}V_x + \mathbf{j}V_y + \mathbf{k}V_z$,

$\mathbf{u} = \mathbf{i}u_x + \mathbf{j}u_y + \mathbf{k}u_z$, and $C_{11}, C_{12}, C_{13}, C_{33}, C_{44}, C_{66} = (C_{11} - C_{12})/2$ the phonon elastic constants, $K_1, K_2, K_3, K_4$ the phason elastic constants, $R$ the phonon-phason coupling constant, $\eta$ the fluid dynamic viscosity, and $\Gamma_u, \Gamma_w$ the phonon and phason dissipation coefficients, $A, B$ the material constants due to variation of mass density, respectively.

The equations (11) are the final governing equations of dynamics of soft-matter quasicrystals of 10-fold symmetry in three-dimensional case with fields variables $u_x, u_y, u_z, w_x, w_y, V_x, V_y, V_z, \rho$ and $p$, the amount of the field variables is 10, and amount of field equations is 10 too, among them: (11a) is the mass conservation equation, (11b)- (11d) the momentum conservation equations, (11e)- (11g) the equations of motion of phonons due to the symmetry breaking, (11h) and (11i) the phason dissipation equations, and (11j) the equation of state, respectively. The equations are consistent to be mathematical solvability, if there is lack of the equation of state, the equation system is not closed, and has no meaning mathematically and physically. This shows the equation of state is necessary.

These equations reveal the nature of wave propagation of fields $\mathbf{u}$ and $\mathbf{V}$ with phonon wave speeds $c_1 = \sqrt{\frac{2A + C_{11} - B}{\rho}}$, $c_2 = c_3 = \sqrt{\frac{C_{11} - C_{12}}{2\rho}}$ and fluid phonon wave speed $c_4 = \sqrt{\left(\frac{\partial p}{\partial \rho}\right)_s}$ and the nature of the diffusion of field $\mathbf{w}$ with diffusive coefficient $D = \frac{1}{\Gamma_w}$ from the view point of hydrodynamics.

## 6. Solving procedure and some preliminary results

The variables $u_x, u_y, u_z, w_x, w_y, V_x, V_y, V_z$ represent the fields of corresponding elementary excitations, which with fluid pressure $p$ and mass density $\rho$ follow the equations (7), (9) and (11) for different soft-matter quasicrystal systems, to determine these hydrodynamic quantities one must solve the equations under appropriate initial and boundary conditions.

The solving of equations (7) (or (9), or (11)) coupled by initial and boundary conditions is very difficult, and the analytic solution might be an exceptional case[15]. The computation is very complex and can be done only by numerical methods. We used finite difference method. In our practice, the computation is stable, which shows the equations and the formulation are correct. The preliminary results exhibit a large distinction of dynamic behaviour between soft-matter and solid quasicrystals, for example, for soft-matter quasicrystals the compressibility $\frac{\delta \rho}{\rho_0} \sim 10^{-4} - 10^{-3}$ [16] by numerical solutions, while for solid quasicrystals $\frac{\delta\rho}{\rho_0} \sim 10^{-13}$ in Ref [17], meanwhile, for soft-matter quasicrystals the ratio between fluid stress over elastic stress $\frac{p_{ij}}{\sigma_{ij}} \sim 1$ are shown in Refs [18], and the ratio between solid viscous stress over elastic stress $\frac{\sigma'_{ij}}{\sigma_{ij}} \sim 10^{-15}$ for solid quasicrystals is demonstrated in Ref [17].

As the solutions require the assistance of mathematical physics and computational physics, and need a large volume for presentation, many details have not included due to space limitations.

**7. Conclusion and Discussion**

The three-dimensional equation systems of generalized dynamics of 8-, 10- and 12-fold symmetry two-dimensional quasicrystals in soft matter were derived. This improves our work previously for planar field.

The solutions of the initial- and boundary-value problems of these nonlinear partial differential equations have provided fruitful results describing mass distribution, deformation and motion of soft-matter quasicrystals, which present great differences in behaviour physically to those of solid quasicrystals.

In the theoretical system, the present dynamics is the heritage and development of the hydrodynamics of solid quasicrystals by Lubensky et al. [8,9].

This article reports only on the three-dimensional version of field equations for the first kind of two-dimensional soft-matter quasicrystals, and it does not concern the second kind of two-dimensional soft-matter quasicrystals, which are more complex, will be discussed later.

More recently Ref [19] reported a molecular dynamics modelling on possible quasicrystals of 12-fold symmetry in smectic B liquid crystals, the work pointed out there might be a problem on the existence of phason degrees of freedom based on the observation in the modeling. This is an interesting topic. In our practice the effect of phason degrees of freedom is very weak due to the decoupling between phonons and phasons in 12-fold symmetry quasicrystals based on the computation with constitutive equations (6).

**Acknowledgements** The authors thank the National Natural Science Foundation of China for the support through Grant 11272053; also thank to Professors T C Lubensky at the University of Pennsylvania, USA; Stephen Z D Cheng at the University of Akron, USA; H H Wensink at the Utrecht University in The Netherlands for helpful discussion.

**References**
[1]Fan T Y, Equation systems of generalized hydrodynamics for soft-matter quasicrystals, *Applied Mathematics and Mechanics*, 2016, **37**, 331-344, in Chinese.
[2]Fan T Y, Generalized dynamics for second Kind of soft-matter quasicrystals, *Applied Mathematics and Mechanics*,2017, **38**,189-199, in Chinese.

[3] Zeng X and Ungar G et al., Supramolecular dendritic liquid quasicrystals, *Nature*, 2004, **428**, 157-159.

[4] A. Takano et al., A mesoscopic Archimedean tiling having a new complexity in an ABC star polymer, *J Polym Sci Pol Phys*, 2005, **43**(18), 2427-2432.

[5] Talapin V D and Shevechenko E V et al., Quasicrystalline order in self-assembled binary nanoparticle superlattices, *Nature*, 2009, **461**, 964-9671.

[6] Fischer S and Exner A, et al., Colloidal quasicrystals with 12-fold and 18-fold diffraction symmetry, *Proc Nat Ac Sci*, 2011, **108**, 1810-1814.

[7] Yue K and Huang M J et al., Geometry induced sequence of nanoscale Frank-Kasper and quasicrystal mesophases in giant surfactants. *Proc Nat Ac Sci*, 2016, **113**(50),1932-1940.

[8] Lubensky T C, Ramaswamy S and Toner J, Hydrodynamics of icosahedral quasicrystals, *Phys Rev B*, 1985, **32**(1), 7444-7452.

[9] Lubensky T C, Symmetry, Elasticity, and Hydrodynamics in Quasiperiodic Structures, Introduction to Quasicrystals, ed Jaric V M, 1988, 199-280, Boston, Academic Press,.

[10] Wensink H H, Equation of state of a dense columnar liquid crystal, *Phys Rev Lett*, 2004, **93**, 157801.

[11] Dzyaloshinskii I E, Volovick G E, Poisson brackets in condensed matter physics, *Ann Phys (NY)*, 1980, **125**(1), 67-97.

[12] Hu C Z, Wang R H and Ding D H, Symmetry groups, physical property tensors, elasticity and dislocations in quasicrystals, *Rep. Prog. Phys.*,2000, **63**, 1-39.

[13] Fan T Y, Mathematical Theory of Elasticity of Quasicrystals and Its Applications, 2010 1st edition, 2016 2nd edition , Beijing, Science Press / Heidelberg, Springer-Verlag.

[14] Fan T Y, Mathematical Theory of Elasticity and Relevant Topics of Solid and Soft-Matter Quasicrystals and Its Applications, 2014, Beijing, Beijing Institute of Technology Press, in Chinese.

[15] Li X F and Fan T Y, Dislocations in the second kind two-dimensional quasicrystals of soft matter, *Physica B*, 2016, **52**, 175-180.

[16] Cheng H ,Fan T Y and Wei H, Impact tension of specimen of 5- and 10-fold symmetry soft matter quasicrsyatals ( refer to Fan T Y,Generalized Dynamics of Soft-Matter Quasicrystals---Mathematical Models and Solutions, 2017, Beijing ,Beijing Institute of Technology Press / Heidelberg ,Springer-Verlag.)

[17] Cheng H, Fan T Y and Wei H, Solutions for hydrodynamics of solid quasicrystals with 5- and 10-fold symmetry, *Appl Math Mech*, 2016, **37**(10), 1393-1404.

[18] Cheng H and Fan T Y, Flow of soft-matter quasicrystals past a circular cylinder, (refer to Fan T Y, Generalized Dynamics of Soft-Matter Quasicrystals---Mathematical Models and Solutions, 2017, Beijing, Beijing Institute of Technology Press / Heidelberg , Springer-Verlag.)

[19] Metere A, Oleynikov P, Dzugutov M and Lidin S, A smectic quasicrystal, *Soft Matter*, 2016, **12**, 8869-8876.